\documentclass[usenatbib, useAMS]{mn2e}
\usepackage{times}
\usepackage[T1]{fontenc}
\usepackage{inconsolata}
\usepackage{bm}
\usepackage{hyperref}
\usepackage{graphicx}
\usepackage{myaasmacros}
\usepackage{enumerate}
\usepackage[english]{babel}
\usepackage{mathptmx}
\usepackage{tabularx}
\usepackage{contour}

\newcommand{\Msol}{\mathrm{M}_{\odot}}
\newcommand{\kpc}{\mathrm{kpc}}
\newcommand{\Mpc}{\mathrm{Mpc}}

\newcommand{\kms}{\mathrm{km\,s^{-1}}}

\def\lsim{\mathrel{\rlap{\lower3pt\hbox{$\sim$}}
    \raise1pt\hbox{$<$}}}                
\def\gsim{\mathrel{\rlap{\lower3pt\hbox{$\sim$}}
    \raise1pt\hbox{$>$}}}                
\newcommand{\pc}{\mathrm{pc}}
\newcommand{\cm}{\mathrm{cm}}
\newcommand{\Myr}{\mathrm{Myr}}

\newcommand{\Gyr}{\mathrm{Gyr}}
\newcommand{\ergs}{\mathrm{ergs}}
\newcommand{\vmax}{v_{\mathrm{max}}}
\newcommand{\yr}{\mathrm{yr}}
\newcommand{\K}{\mathrm{K}}
\newcommand{\Mvir}{M_{\mathrm{vir}}}
\renewcommand{\vec}[1]{\contour[1]{black}{$#1$}}
\renewcommand{\vec}[1]{\bm{#1}}
\newcommand{\changed}[1]{{\color{red}{#1}}}
\renewcommand{\changed}[1]{#1}

\begin{document}
\pubyear{2016} \title[How to quench a galaxy]{How to quench a galaxy}
\author[A. Pontzen et al]{Andrew Pontzen$^1$, Michael Tremmel$^2$,
  Nina Roth$^1$, Hiranya V. Peiris$^1$, \and Am\'elie Saintonge$^1$,
  Marta Volonteri$^3$, Tom Quinn$^2$, Fabio Governato$^2$
  \\
  $^1$ {Department of Physics and Astronomy, University College
    London,
    London WC1E 6BT, UK} \\
  $^2$ {Astronomy Department, University of Washington, Seattle, WA
    98195, US} \\
  $^3$ {Institut d'Astrophysique de Paris, 98 bis bd Arago, 75014
    Paris, France} }

\date{ Received ---; published---. }
\maketitle

\begin{abstract}
  We show how the interplay between active galactic nuclei (AGN) and
  merger history determines whether a galaxy quenches star formation
  at high redshift.  We first simulate, in a full cosmological
  context, a galaxy of total dynamical mass $\Mvir = 10^{12}\,\Msol$
  at $z=2$. Then we systematically alter the accretion history of the
  galaxy by minimally changing the linear overdensity in the initial
  conditions. This ``genetic modification'' approach allows the
  generation of three sets of $\Lambda$CDM initial conditions leading
  to maximum merger ratios of 1:10, 1:5 and 2:3 respectively. The
  changes leave the final halo mass, large scale structure and local
  environment unchanged, providing a controlled numerical
  experiment. Interaction between the AGN physics and mergers in the
  three cases lead respectively to a star-forming,
  temporarily-quenched and permanently-quenched galaxy. However the
  differences do not primarily lie in the black hole accretion rates,
  but in the kinetic effects of the merger: the galaxy is resilient
  against AGN feedback unless its gaseous disk is first
  disrupted. Typical accretion rates are comparable in the three
  cases, falling below $0.1\,\Msol\,\yr^{-1}$, equivalent to around
  $2\%$ of the Eddington rate or $10^{-3}$ times the pre-quenching
  star formation rate, in agreement with observations. This low level
  of black hole accretion can be sustained even when there is
  insufficient dense cold gas for star formation. Conversely,
  supernova feedback is too distributed to generate outflows in
  high-mass systems, and cannot maintain quenching over periods longer
  than the halo gas cooling time.\vspace{0.5cm}
\end{abstract}

\section{Introduction}

Reproducing the population of galaxies observed in the Universe within
a $\Lambda$CDM cosmological paradigm requires significant energetic
feedback \citep{WhiteFrenk91Review} to prevent over-cooling and
excessive star formation. In fact star formation must be inefficient
in both low and high mass dark matter halos, with a peak efficiency
corresponding approximately to the luminosity function turnover
$L_\star$
\citep[][]{Bower06AGNQuenching,Guo10AbundanceMatching,Moster13AbundanceMatch,Behroozi13AbundanceMatch}.
Regulation of galaxies with luminosities $L<L_{\star}$ can be
attributed to ultraviolet background radiation slowing the accretion
of gas into shallow potential wells
\citep[][]{Efstathiou92UV,Bullock00UV,Somerville02UV}, and
subsequently energy input from young stars
\cite[][]{WhiteRees78,Cole94Semianalytic,EfstathiouSNfeedback,Governato07,Pontzen08DLAs}. The
relatively shallow potential wells mean that the typical speeds
achieved by supernova- and radiation-driven galactic winds can exceed
the escape velocity \citep{MacLow99outflows,Christensen15outflows}.
However many lines of reasoning suggest that these processes must
become increasingly ineffective as dynamical masses approach
$10^{12}\,\Msol$
\citep[][]{Benson03,Bower06AGNQuenching,Hopkins14FIRE,KellerWadsley16}.

Accounting for reduced star formation at high masses has proved more
contentious. Early scaling arguments and semi-analytic models
suggested that long cooling times associated with the raised virial
shock temperature could be a simple explanation
\citep{Binney77,ReesOstriker77,WhiteRees78,Kauffmann93semianalytic,Cole94Semianalytic,Somerville99semianalytic}. Unfortunately
the resulting suppression is only effective for low baryon fractions,
making it insufficient to account for observed luminosity functions
given today's concordance $\Lambda$CDM parameters \citep{Benson03}. A
related possibility is offered by starvation of galaxies simply
because accretion slows down at late times
\citep{Feldman2015AccretionQuenching,Feldman2016AccretionQuenching}. However
the host halo masses of red galaxies are observed to be systematically
lower than those of star-forming galaxies, suggesting that accretion
continues after star formation shuts down
\citep{Mandelbaum16LensingMasses}.  Semi-analytic models also suggest
that without additional mechanisms, it is very difficult to
quantitatively explain the division of the population into actively
star-forming disks and ``quenched'', red ellipticals
\citep{Baldry04Bimodality,Bower06AGNQuenching}. This division appears
to be in place even at high redshift
\citep{Ilbert10COSMOSbimodality,Brammer11bimodality}.

Environmental effects, especially in high-density group and cluster
regions can strip an infalling galaxy of its gas reservoir which leads
to a more natural bimodality
\citep[e.g.][]{QuilisMoore2000EnvQuenching,Gomez03SFRvsEnv,Boselli06,vdBosch08SatelliteQuenching,McCarthy08SatelliteQuenching,Bahe15GroupQuenching}. However,
quenching also occurs outside of these environments
\citep{vdBosch08SatelliteQuenching,Guo09GroupGalaxyProperties,Peng10Quenching,Wijesinghe12gamaEnvironmentSFR}.
We are therefore led to re-examine the role of feedback. One route to
increasing the efficacy of stellar feedback at intermediate and high
masses is to induce intense, compact starbursts through major mergers
or violent disk instabilities
\citep{DekelBurkert14,Ceverino15BulgeFormation,Zolotov15Quenching,Tacchella16Compaction}. However
quenching in this scenario remains fairly gradual, with star formation
declining over several gigayears \citep{Zolotov15Quenching}, whereas a
number of lines of evidence suggest that a significant fraction of early type galaxies are
formed by much more rapid quenching
\citep[e.g.][]{Thomas05,Schawinski14GreenValley,Belli15,Barro16RapidQuenching}.
Moreover, in this picture, quenching can be maintained over long time
periods only in high-mass galaxies where the virial shock prevents
rapid cooling of material accreted after the starburst event
\citep{WhiteFrenk91Review,BirboimDekel03}. By definition, quenching
star formation for longer than the cooling time of halo gas requires
an energy source other than stellar feedback.

Many studies suggest that the crucial input comes from active galactic
nuclei (AGN) powered by central supermassive black holes; see, for example,
\cite{tdMatteo2005Quasars,Hopkins05BHgalaxyMergers,Bower06AGNQuenching,Croton06,Sijacki07AGNfeedback,diMatteo2008QuasarsCosmo,Cattaneo09BHreview,Johansson09,Fabian12BHreview,Dubois13AGNGalFormation,Dubois16}
and references therein.  AGN drive rapid outflows which can be
directly observed in post-starburst galaxies
\citep[e.g.][]{Tremonti07OutflowsPostStarburst,RupkeVeilleux11},
suggesting that black holes are able to suppress star formation by
removing the supply of gas.  Adding weight to the connection between a
galaxy and its central black hole (BH) is the strong observed correlation
between BH mass and stellar mass in the bulge
\citep{Cattaneo09BHreview,Volonteri12BHreview,Kormendy13BH_annrev},
which can be interpreted either as evidence that the BH and SF are fed
from the same supply of cold gas; that feedback from BH regulates the
star formation rate; or even, for low-mass galaxies, that feedback
from SF regulates the BH accretion \citep{Dubois15BHSNinteraction}. It
is possible that the true explanation is a combination of all three
effects.  But whatever the regulatory role that BHs play,
observations also show that AGN activity is common in highly
star-forming galaxies
\citep[e.g.][]{Nandra07AGNhostgalaxies,Simmons12AGN,Mullaney2012,Rosario12SFRofAGN,Rosario13AGNandSFcorrelate,ForsterSchreiber14SinsOutflows,Mullaney15SFRofAGN,Carniani16QuasarsDontAlwaysQuench},
suggesting that the precise role of the BH is strongly
dependent on other, unidentified factors.

In this paper, we will identify those factors by studying the
interaction between AGN feedback and mergers in a realistic
cosmological environment, discussing how it can lead to quenching that
is maintained for periods longer than the halo cooling time.  To
address this question we require an approach that offers control over
environment, accretion history, and feedback models.  The basis for
our study is a $z=2$ galaxy with a dynamical mass of
$10^{12}\,\Msol$, simulated using the \textsc{ChaNGa} code
\citep{Changa15}.  Uniquely, we are able to modify the accretion
history of the system by making minimal modifications to the
large-scale structure. This ability to ``genetically modify'' the
system that we study, while keeping the cosmological conditions
consistent with the $\Lambda$CDM inflationary scenario, is introduced
by \cite{Roth15GM}. Our main aim here is to study how the BHs
respond to changing the significance of the most major merger. The
combination of being able to control feedback and history with a fixed
local environment (including the precise directions of the filaments
that feed cool gas to the galaxy) allows us to isolate and identify
the conditions required for quenching. In particular, all simulations are run
first with and then without BHs to quantify the effect of the AGN
feedback.

This paper is structured as follows. In Sec. \ref{sec:simulations}
we describe the simulations in more detail. The results are discussed
in Sec. \ref{sec:results}. Finally we conclude in Sec.
\ref{sec:concl-disc}.

\section{Simulations}\label{sec:simulations}

\subsection{The genetic modification approach}\label{sec:genet-modif-appr}

{The morphology and colour of a galaxy is determined jointly
  by its mass, environment and history in tandem with internal
  feedback processes. Idealised
  simulations allow one to study the collision of two isolated galaxies, and
  consequently form part of the existing evidence that black holes
  and mergers have a role to play in quenching
  \cite[e.g.][]{tdMatteo2005Quasars,Johansson09}. However, such a
  computational approach lacks a cosmological environment and so
  neglects the inflowing gas filaments that provide fresh material for
  star formation; the results show whether feedback is able to disrupt
  the existing cold gas content but do not address the galaxy's
  evolution after this point. Conversely, traditional cosmological
  simulations include the interaction between a black hole
  and continuing gas accretion
  \changed{\citep[e.g.][]{Dubois13AGNGalFormation,Choi15}}, but do not allow one
  to build a predictive understanding of the dependence on merger history.

  In this work we combine the best features of the two approaches: we
  study the interplay between mergers and BH feedback by using} the
genetic modification \citep[GM; see][]{Roth15GM} approach to generate
a series of closely-related initial conditions which lead to different
accretion histories. Our three initial conditions share the same
large-scale structure and power spectrum, allowing us to make a
controlled study while maintaining features such as the filaments
along which gas streams.

We start with an unmodified reference zoom cosmological simulation of
a galaxy of virial mass $\Mvir \simeq 10^{12}\,\Msol$ (see Sec.
\ref{sec:running-simulations} for cosmological and numerical
parameters). The linear overdensity field on the initial grid is
denoted by the vector $\vec{\delta}_0$; each particle in the
simulation then maps to a particular element of this initial
vector. To increase (or decrease) the significance of a particular
merger, we identify the particles of the infalling substructure and
increase (or decrease) their mean overdensity in the initial linear
vector. To compensate the final mass, the mean overdensity of the
elements mapping to the final halo at $z=2$ must be kept fixed.

It is not possible to simply modify the $\vec{\delta}_0$ linear
overdensity vector by hand, since this would typically make the
resulting field an extremely unlikely draw from the Gaussian random
ensemble. The \cite{Roth15GM} approach chooses a modified field
$\vec{\delta}_1$ that is as close as possible to the original
$\vec{\delta}_0$ in the sense of minimising
$(\vec{\delta}_1-\vec{\delta}_0)^{\dagger} \mathbf{C}^{-1}
(\vec{\delta}_1 - \vec{\delta}_0)$ while also satisfying the
imposed overdensity modifications\footnote{We implement the field
  changes by extending the code described by \cite{Roth15GM} to allow
  the spatially-varying resolution required for zoom simulations. The
  mathematical formalism does not change; the technical implementation
  will be described in a forthcoming paper. }. Here, $\mathbf{C}$ is
the covariance matrix generated from the $\Lambda$CDM power spectrum.

For the purposes of this work, we generated three versions of a galaxy
in the same large scale structure but with differing merger
histories. The specific modifications are described further in Sec.
\ref{sec:merger-histories}. We verified that our
constraints lead to a field that is consistent with being drawn from a
$\Lambda$CDM power spectrum by calculating that $\chi^2 \equiv
\vec{\delta}^{\dagger}_1 \mathbf{C}^{-1} \vec{\delta}_1$ remains
close to $N$, the number of particles (i.e. degrees of freedom) in the
simulation. In the modifications described below, $\Delta \chi^2 = \pm
2$, which is a small perturbation around the $N \simeq 1.9 \times
10^7$ degrees of freedom in the box. Since the modifications are
concentrated in the zoom region, one might instead compare to $N
\simeq 1.7 \times 10^6$ in the zoom region. Seen either way, the
modifications produce a field that is strongly consistent with the
$\Lambda$CDM power spectrum.

\subsection{Running the simulations}\label{sec:running-simulations}

The simulations are performed using the task-based parallelised
Tree-SPH code \textsc{ChaNGa} which uses a geometric mean-density SPH
interpolation approach, reducing artificial surface tension to better
resolve fluid instabilities \citep{Changa15}. 
{The physics implemented includes star formation, supernova feedback
and metal cooling. We also include a new prescription for BH
formation, accretion and feedback. One key recommendation from
multiscale studies of accretion on multiple scales
\citep[e.g.][]{Hopkins10AngMomBH,Hopkins11AngMomBH} is that angular
momentum support plays a critical role in determining BH growth in disk
galaxies \citep[see
also][]{RosasGuevara15AngMomLimitedAccretion,AnglesAlcazar16BHAngMom}.
Our accretion model is therefore based on a modified Bondi-Hoyle
formula that accounts for the rotational support of gas on resolved
scales without requiring additional sub-grid assumptions
\citep{Tremmel16Romulus}. Additionally our approach includes a
dynamical friction prescription \citep{Tremmel15BHDF} that produces
physically-motivated, resolution-independent predictions for BH
trajectories and mergers. 

Feedback from stellar winds, Type II and Ia supernovae, and black
holes is deposited thermally in the SPH kernel surrounding the
source. In contrast to the dynamics, the feedback is therefore
explicitly resolution-dependent, reflecting that the physical
processes underlying feedback are always unresolved and must be
represented by an effective model. Accordingly we are left with a
handful of numerical parameters; these are calibrated against the
$z=0$ luminosity function and BH -- bulge-mass relation. The tuning
procedure, and its verification using a $(25\,\Mpc)^3$ uniform
resolution volume simulation known as \textsc{Romulus25}, is described
by \cite{Tremmel16Romulus}. At present we have only calibrated the new
feedback models at a single resolution, and we therefore adopt this
resolution for the present work (see below). To isolate the effect of
the BH feedback, we run each of our initial conditions twice; first
including all forms of feedback (which we will refer to as a
``BH+SNe'' run), then again including only stellar feedback
(``SNe-only'').

}

We assume a {\it Planck} $\Lambda$CDM cosmology throughout
\citep[based on Table 3, column 4 of][]{Planck2015parameters}.  We
start by simulating a $z \simeq 2$ field galaxy with a quiet evolution
history.  In preparation, we performed a $50\,\Mpc$ uniform volume
dark-matter-only simulation, from which we selected halos with virial
masses at $z=2$ in the range
$0.9 \times 10^{12}<\Mvir/\Msol<1.1\times 10^{12}$ with the intention
of obtaining a galaxy of stellar mass
$M_{\star} \simeq 5 \times 10^{10} \Msol$. We eliminated objects
within a comoving megaparsec of a more massive halo, leaving four
candidates from which we picked the object with the smoothest
accretion history. The biggest event in the halo's history to $z=2.0$
is a 1:5 merger at $1.7\,\Gyr$
({$z=3.7$}).\footnote{{Here, and throughout, we define
    merger ratios using the mass of bound dark particles as
    reported by Amiga's Halo Finder \citep[AHF,][]{AHF} at the time that the
    infalling structure's mass peaks.}} At $z=2.0$, the
DM-only mass is $9.7\times 10^{11}\,\Msol$; by $z=0.0$ this has grown
to $3.3\times 10^{12}\,\Msol$.

We refined the density field around the target halo to generate
zoom initial conditions \citep{KatzWhite93} and re-ran
the simulation adding baryons. Our final simulation suite has a
resolution of $M_{\mathrm{p,gas}} = 2.1 \times 10^{5}\,\Msol$ and
$M_{\mathrm{p,dm}} = 1.4 \times 10^{5}\,\Msol$. (To achieve these
similar particle masses, which is desirable to reduce noise in the potential near the
centres of galaxies, the dark matter field is realised at twice
the linear resolution of the gas field.) The Plummer-equivalent
softening in the high-resolution region is $250\,\pc$.

\begin{figure*}
\includegraphics[width=1.0\textwidth]{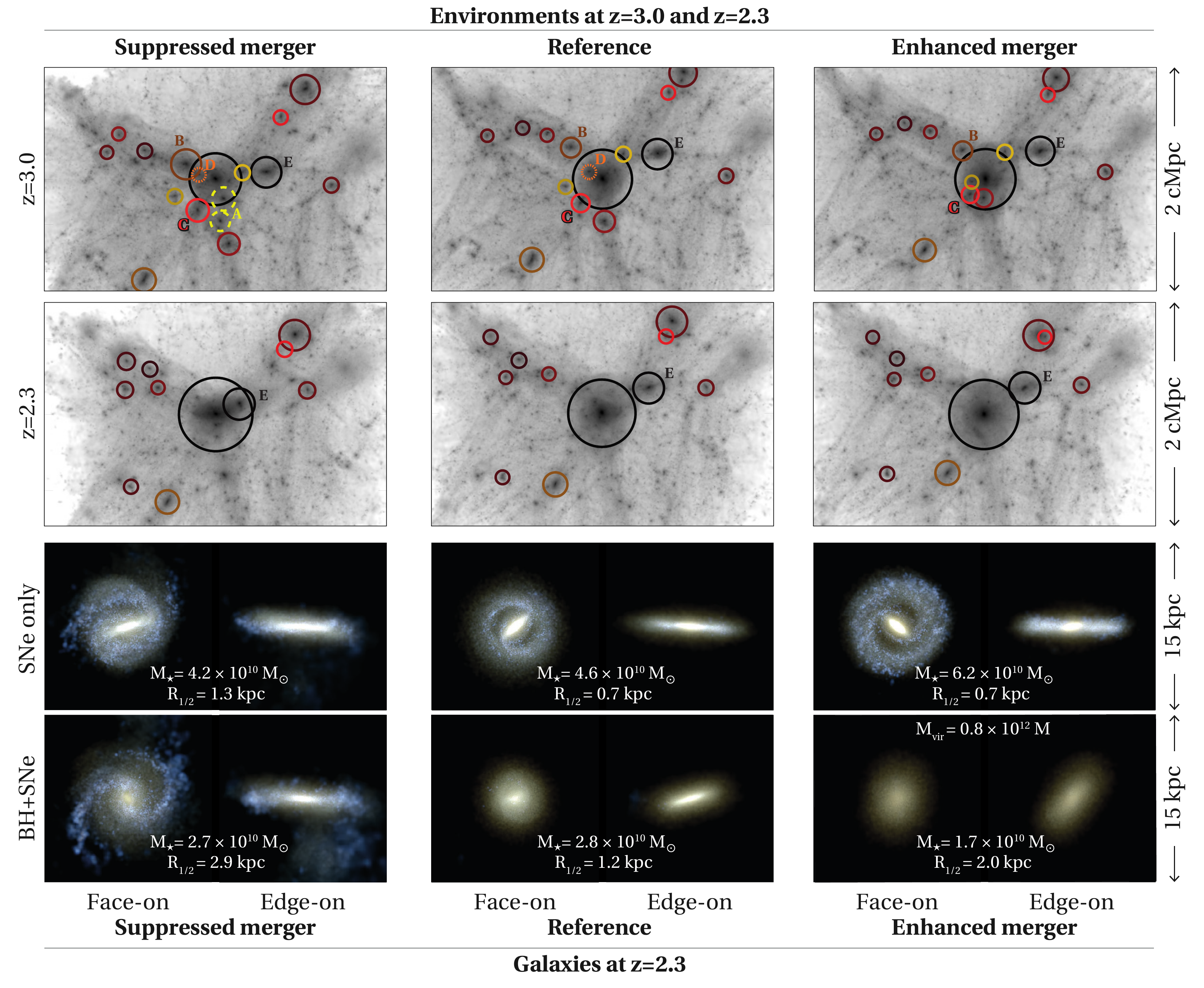}
\caption{The genetic modification approach minimally changes the large
  scale structure while systematically altering the mass accretion
  history, to provide a controlled test of galaxy quenching.  {\it
    (Upper two rows)} the projected density of dark matter in the high
  resolution region surrounding the SNe+BH galaxies at $z=3.0$ and
  $z=2.3$ respectively; the images are scaled to show dark matter
  column densities of $10^{19}$ to $10^{23}\,m_p\,\cm^{-2}$, where
  $m_p$ is the proton mass. The most massive individual halos are
  identified by circles at the virial radii; halos are consistently
  labelled between the panels so that individual structures can be
  identified. Dotted/dashed halos have already merged into the main
  halo in the reference (A) and enhanced-merger (A and D)
  simulations. The significance of the other labelled structures is
  discussed in Sec. \ref{sec:merger-histories}.  {\it (Lower two
    rows)} portraits of the galaxy in IVU rest-frame colours at
  $z=2.3$ ($t=2.8\,\Gyr$, i.e. a lookback time of $11.0\,\Gyr$). The
  images are scaled to cover a large dynamic range of $16$ to
  $22\,\mathrm{mag}\,\mathrm{arcsec}^{-2}$. The upper of the two rows
  shows the SNe-only simulation suite; the galaxy forms a bright
  central bar/bulge. As our merger is made more significant and moved
  to earlier times, this becomes more concentrated. The bottom row
  depicts galaxies from the BH+SNe suite; the highly concentrated bars
  are absent, and the galaxy becomes increasingly quiescent, red and
  spheroidal as the merger ratio increases from left to
  right. }\label{fig:portraits}
\end{figure*}

\subsection{The merger histories}\label{sec:merger-histories}

\begin{figure}
\includegraphics[width=0.48\textwidth]{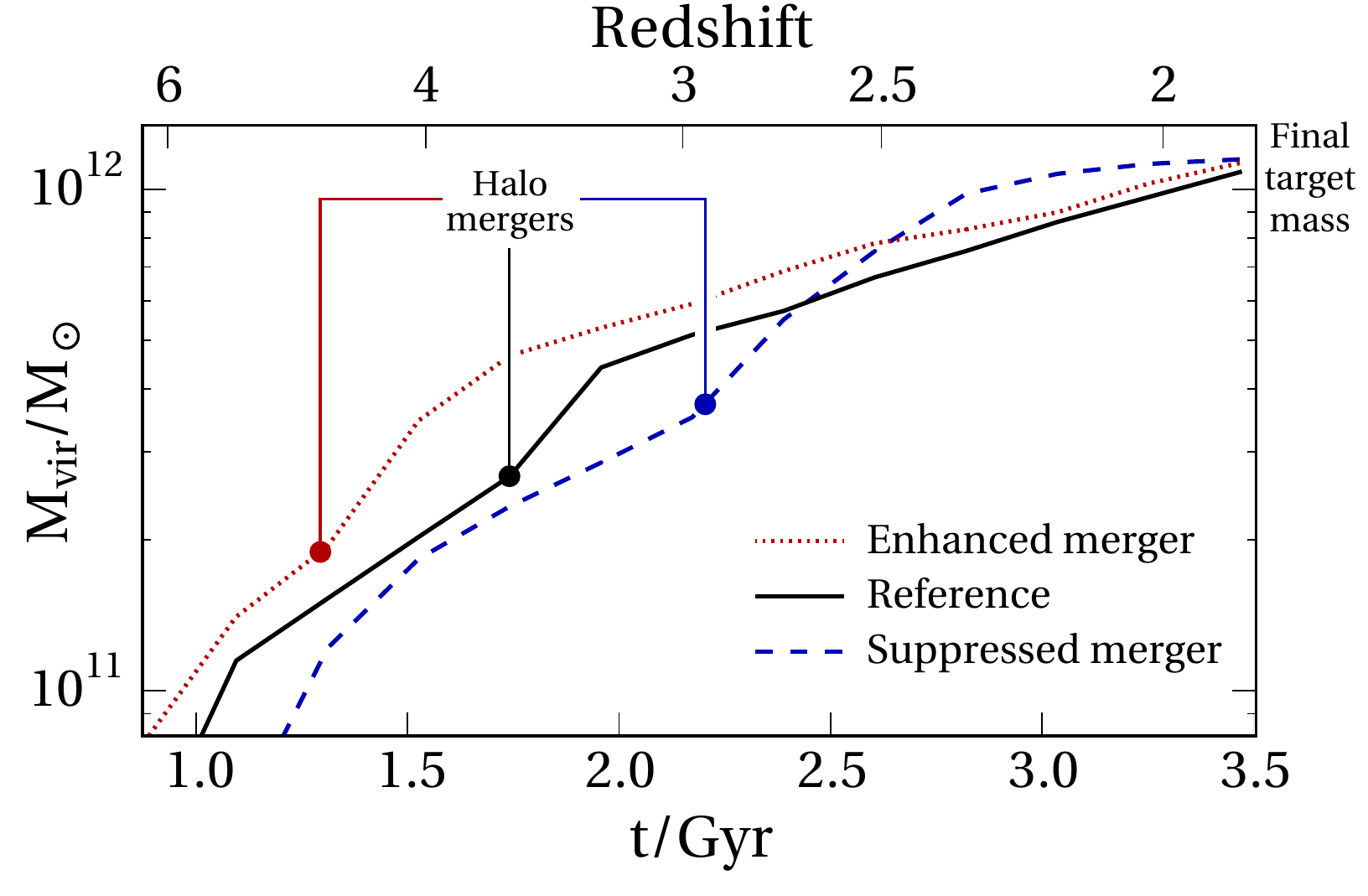}
\caption{The virial mass over time of our three scenarios, showing the
  effect of the GM alterations.  The black, red-dotted and blue-dashed
  lines refer to the reference, enhanced-merger and suppressed-merger
  runs in the SNe-only suite respectively. The build-up of mass
  differs, but by construction the final mass is fixed at the end of
  the simulations, $z=1.8$. }\label{fig:mvir-vs-time}
\end{figure}

Using the GM approach (Sec. \ref{sec:genet-modif-appr}), we
generate three different initial conditions; after running the
simulations, we can study the corresponding three merger scenarios
for our numerical experiment. The halo mass history is shown in
Fig.~\ref{fig:mvir-vs-time}. In summary:
\begin{itemize}
\item The original (``reference'') system's largest merger prior to
  ${z=2}$ has a mass ratio of 1:5 at $z=4.0$ ($t=1.5\,\Gyr$). Note that the
  galaxies merge slightly later than the parent halos, at $z=3.2$
  ($t=2.0\,\Gyr$).
\item Our first genetic modification is designed to generate a more
  significant merger by increasing the mass of the infalling
  object. This changed mass also causes the structure to fall in
  somewhat earlier; specifically we obtain a 2:3 merger at $z=4.6$
  ($t=1.3\,\Gyr$; for the galaxies, $z=3.7$ and $t=1.7\,\Gyr$). We
  call this the {\it enhanced merger} simulation.
\item Our second modification is designed to generate a smaller (but
  therefore later) merger.  We found that due to the fixing of the
  final virial mass, it is accompanied by a rapid series of
  individually small accretion events that rapidly build mass
  (described below) around $z=2.3$ ($t=2.8\,\Gyr$; for the galaxies,
  $z=2.5$ and $t=2.6\,\Gyr$). We call this the {\it suppressed merger}
  simulation.
\end{itemize}

The top row of panels in Fig. \ref{fig:portraits} show the situation
at $z=3.0$ ($t=2.2\,\Gyr$). At this time, the target halo has already
merged in the reference and enhanced-merger scenarios (centre and
right columns respectively). The large scale structure is therefore
close to identical in the two cases, with only small changes in the
positions of some individual objects.  However at this time in the
suppressed-merger case (left column), the merger is just taking
place. In fact the target merging substructure has actually been split
into two halos, marked A. The uppermost of these halos is just about
to merge with a ratio of 1:10, while the lower constitutes an
additional minor merger of 1:25 at $z=2.3$ ($t=2.8\,\Gyr$). The
genetic modification, which targets a fixed $z=2$ mass as an
additional constraint, has partially compensated for the lower total
mass present in accreted system A by increasing the mass of the
separate infalling system B. This merges at $z=2.5$ with a ratio 1:10
-- whereas in the reference and enhanced-merger simulations the ratio
for system B is 1:25 and 1:36 respectively. Meanwhile system C has a
ratio of 1:20 and merges at $z=2.5$ in all cases. Accompanied by a
large number of other even smaller substructures such as system D
(1:60), the net effect is that the suppressed-merger halo grows
rapidly around this time but without any major mergers.

In all three cases, system E tidally interacts with the main
halo but is on a tangential trajectory and does not merge in the
lifetime of our simulations. The closest approach occurs at $z\simeq
2.3$ ($t=2.8\,\Gyr$), illustrated in the second row of Fig.
\ref{fig:portraits}.

\section{Results}\label{sec:results}

The lower panels of Fig.~\ref{fig:portraits} show  portraits of the
galaxy in UVI
rest-frame colours at $z=2.3$ in the three SNe-only runs (middle row)
and the corresponding SNe+BH runs (bottom row) with luminosity scaling
between $16$ and $22\,\mathrm{mag}\,\mathrm{arcsec}^{-2}$. Each simulation  is
shown in a face-on and edge-on projection, established according to
the angular momentum of the baryons. The images are not
dust-attenuated.

\begin{figure}
\includegraphics[width=0.48\textwidth]{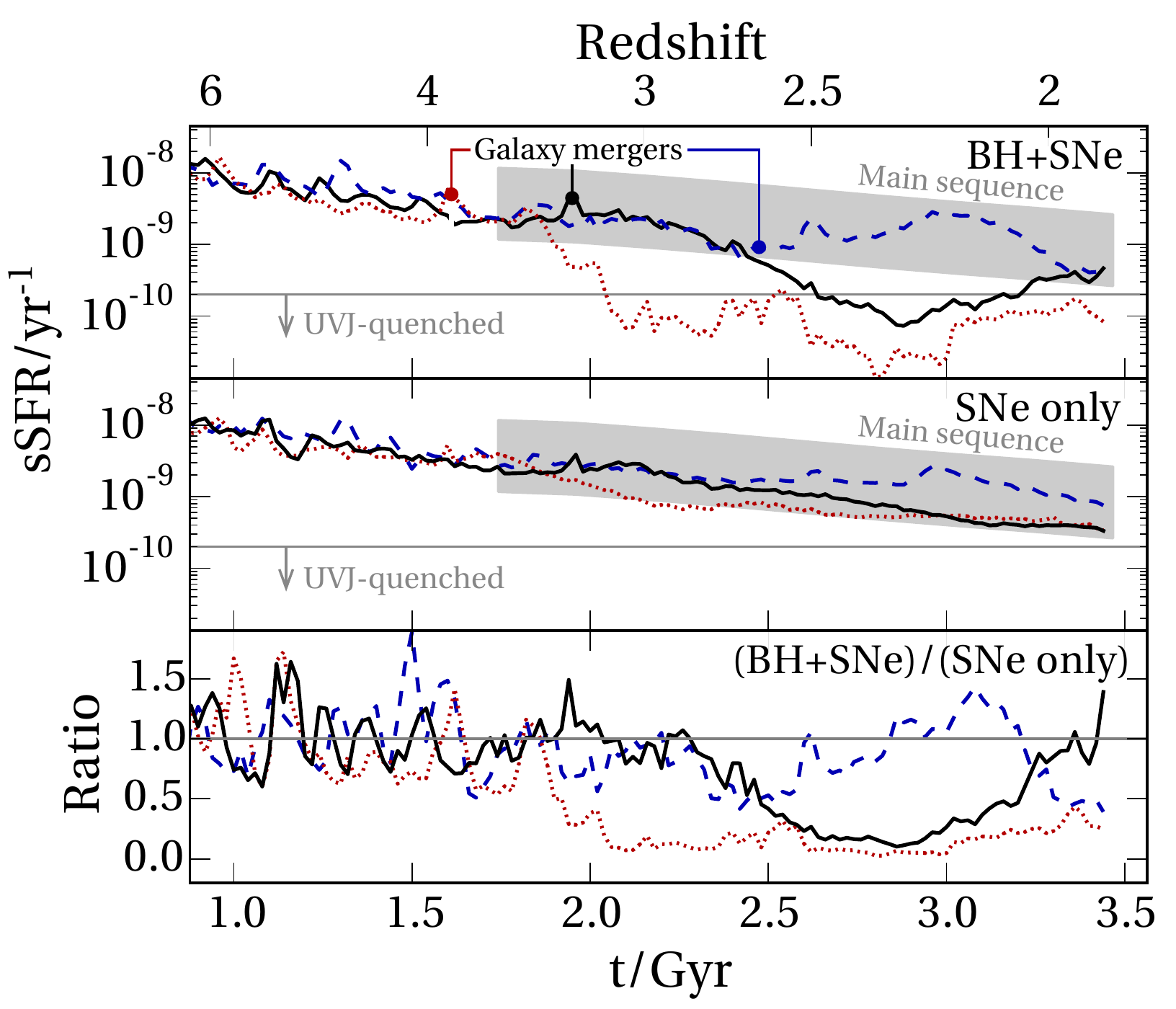}
\caption{The interaction between BH feedback and merger history
  determines whether a galaxy quenches.  {\it Top two panels}: The
  specific star formation rate (sSFR) of our three galaxies in the
  BH+SNe and SNe-only simulation suites respectively. Colours and
  linestyles correspond to those in Fig.~\ref{fig:mvir-vs-time}. In
  the BH+SNe suite, the enhanced-merger case quenches permanently, the
  reference simulation quenches temporarily, and the suppressed-merger
  simulation never quenches. The grey band shows the approximate main
  sequence expectation using the fits from ZFOURGE supplemented by
  far-IR imaging \citep{Tomczak16SFRZFourge} evaluated at the stellar
  mass of the reference galaxy. When galaxies fall below
  $2\times 10^{-10}\,\yr^{-1}$ (horizontal lines), they are defined as
  quenched by the widely-used UVJ observational cut.  {\it Bottom
    panel}: The ratio between the sSFRs of each merger scenario for
  the two different feedbacks, underlining how the relative effect of
  BH feedback is strongly dependent on history.}\label{fig:ssfr}
\end{figure}

In the BH+SNe suite, the original and enhanced-merger scenarios give
rise to a quenched, elliptical galaxy at $z=2.3$. This is confirmed by
analysis of the specific star formation rates (sSFRs, defined by $\mathrm{sSFR}
\equiv {\dot{M}_{\star} / M_{\star}}$) in Fig.
\ref{fig:ssfr}. The first and second panels show the BH+SNe and
SNe-only simulation suites respectively. We also calculate the main
sequence star formation rate $\mathrm{SFR}(M_{\star}, z)$ using the
fitting formula from \cite{Tomczak16SFRZFourge}, itself based on
ZFOURGE\footnote{The FourStar Galaxy Evolution Survey,
  \url{http://zfourge.tamu.edu}} supplemented by far-infrared
observations from {\it Herschel} and {\it Spitzer}. The result is
plotted with $\pm 0.5\,\mathrm{dex}$ scatter as grey bands.
For the input stellar mass $M_{\star}$ we use the reference simulation
(noting that the sensitivity of the sSFR to $M_{\star}$ is relatively
weak). It can be observationally convenient to define a quenched
galaxy by its locus in the rest-frame UVJ colour plane
\citep[e.g.][]{Williams09UVJQuenching,Whitaker11UVJQuenching}; this
results in an upper limit on the specific star formation rate of
approximately $2\times 10^{-10}\,\yr^{-1}$
\citep[e.g.][]{Shards2016Quiescence}, which is shown as a horizontal
line in Fig. \ref{fig:ssfr}.

In the SNe-only suite, the merger history has a relatively small
effect, with star formation remaining active in all three
cases. However in the BH+SNe case (top panel of Fig.  \ref{fig:ssfr}),
the behaviour becomes highly sensitive to history. The remainder of
this Section explores why the BH feedback is sensitive to accretion
history, and especially why it leads to a rapid and long-term
quenching in the enhanced-merger scenario at $t=2.0\,\Gyr$. First, we
will inspect the accretion rates of the BHs
(Sec. \ref{sec:how-quench}) before showing that the coupling rather
than the total amount of energy is the key differentiating factor of
the BH+SNe suite (Sec. \ref{sec:energy-coupling-matters}). Sec.
\ref{sec:sustain-quenching} shows that the BH is required not just to
initiate quenching but to sustain it. However we also show in
Sec. \ref{sec:how-unquench} that the BH actively accretes in the
suppressed-merger scenario; star formation is allowed to continue
because the AGN is kept under control by the dense, rotation-supported
gaseous galaxy
disk. Sections~\ref{sec:temporary-or-permanent}~and~\ref{sec:semi-enhanced-sim}
draw implications for the population of galaxies as a whole. Finally,
in Sec.~\ref{sec:dynamical-effects}, we briefly consider the dynamical
impact of the BH feedback.

\subsection{Black hole accretion rates remain strongly sub-Eddington, even
  during mergers}\label{sec:how-quench}

\begin{figure}
\includegraphics[width=0.48\textwidth]{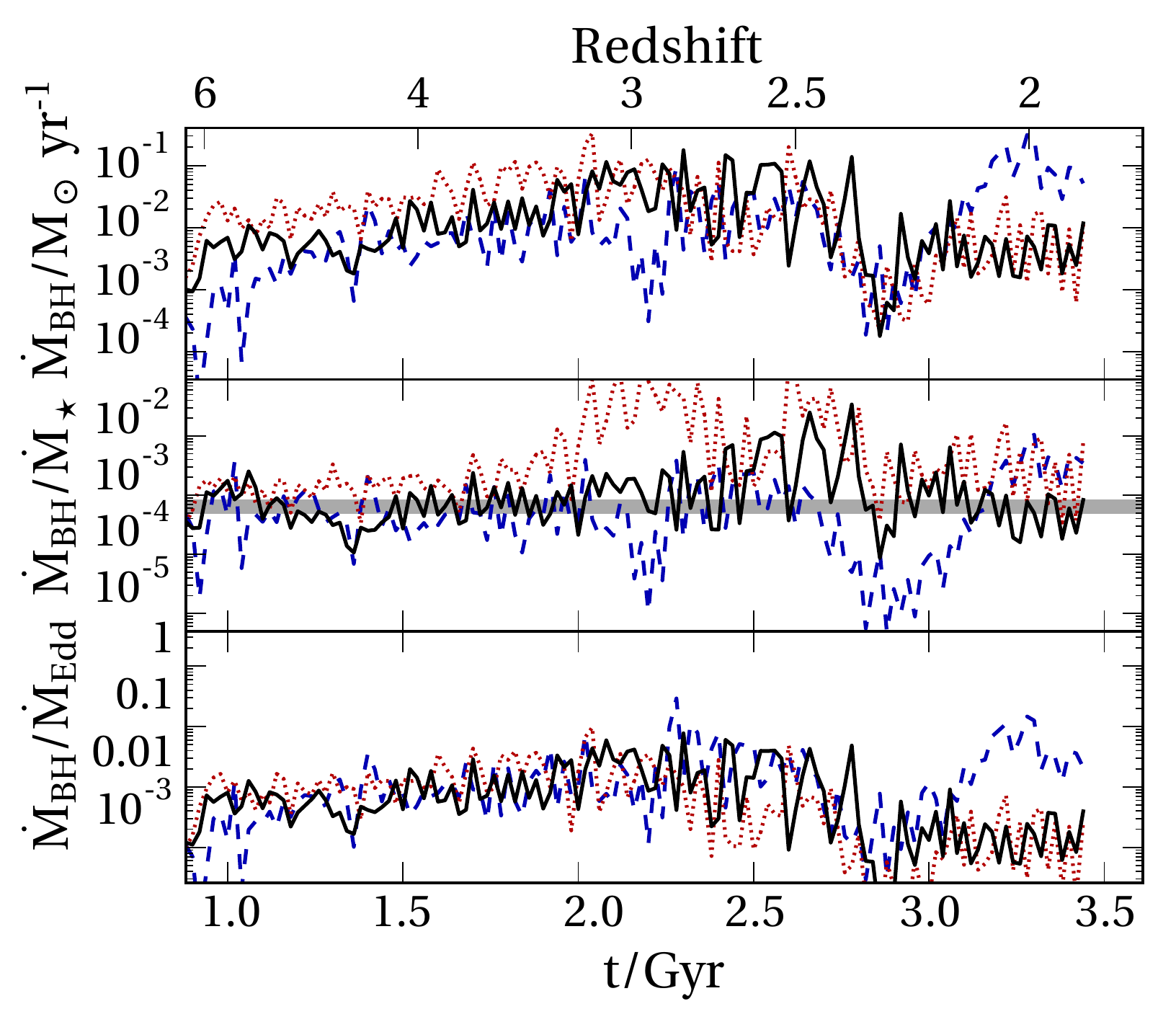}
\caption{The BH feedback operates at a low Eddington ratio, even
  during mergers. From top to bottom, the total BH accretion rate is
  shown as an absolute rate, as a fraction of the star formation rate
  and as a fraction of the Eddington rate. Before quenching, all three
  galaxies closely conform to the mean observational relation from
  \protect\cite{Mullaney2012} (grey band, middle panel). At later
  times, the luminosity of the BH is sustained and therefore increases
  as a fraction of the star formation rate. The tidal interaction with
  halo E (see Fig.~\protect\ref{fig:portraits}) causes a drop in
  accretion rate in all three scenarios at $t \simeq 2.8\,\Gyr$
  followed by a slight rise in the suppressed-merger case, explained
  in the text.}\label{fig:bh-energy}
\end{figure}

Figure~\ref{fig:bh-energy} shows three ways of quantifying the BH
accretion rate over time in the three BH+SNe simulations. As before,
reference, enhanced, and suppressed merger simulations are represented
by black solid, red dotted and blue dashed lines. The top panel shows
the total BH accretion rate $\dot{M}_{BH}$. In our prescription, which
follows the physical orbit of BHs rather than tying them to the halo
centre, there can be more than one BH per halo; at each timestep we
therefore sum over all that are in the major progenitor
halo. Nonetheless, we found that the total mass and the accretion rate
are both strongly dominated by one central BH except for a few
timesteps around mergers.

Note that the drop in accretion rate at $t\simeq 2.8\,\Gyr$ in all
three scenarios is linked to a strong tidal interaction at closest
approach of Halo E (as shown in Fig.  \ref{fig:portraits}). While this
modestly increases the central density, it also increases the
significance of the angular momentum support term
\citep{Tremmel16Romulus}, so that the net effect is an overall
reduction in the accretion rate. After the tidal event, the accretion
rates recover. In the suppressed-merger case, the AGN becomes offset
relative to the galaxy centre by $500\,\pc$, leading to an increased
accretion rate as its orbit decays. The consequences will be
discussed in Sec. \ref{sec:how-unquench}.

The middle panel shows the accretion rate $\dot{M}_{BH}$ normalised to
the star formation rate $\dot{M}_{\star}$. The grey band represents the
observed relationship averaged over many galaxies,
$\langle\dot{M}_{BH} \rangle \simeq (0.6 \pm 0.1) \times 10^{-3}
\langle\dot{M}_{\star}\rangle$
at $\langle z \rangle=2$ \citep{Mullaney2012}. The redshift evolution
of this relationship is extremely weak at least to $z=3$
\citep{Stanley15AGNGalaxyConnection}, and so we have plotted a single
mean relationship. Physically, the relationship is thought to derive
from a common gas infall rate to the BH and stellar components
\citep{Dai15AGNGalaxyConnection,Harris16QSOGalaxyConnection};
individual objects scatter significantly around the mean relation
\citep{Stanley15AGNGalaxyConnection}, since this connection is
imperfect.

The final panel of Fig.~\ref{fig:bh-energy} shows the accretion as a
fraction of the Eddington limiting rate, defined by
$\dot{M}_{\mathrm{Edd}} = 2.2 \times 10^{-8}\, M_{\mathrm{BH}}\,
\mathrm{yr}^{-1}$ given the $10\%$ radiative efficiency assumed by our
BH implementation.  
The accretion rate averages $1.6\%$
of Eddington, even during the merging and quench phases, with only
occasional bursts reaching a maximum of $15\%$. This contrasts with
some earlier works in which an Eddington-limited phase is invoked
\cite[e.g.][]{tdMatteo2005Quasars,Sijacki07AGNfeedback,Sijacki2015IllustrisAGN};
our lowered Eddington fractions are more consistent with constraints
from the observed quasar population
\citep[e.g.][]{Kelly10QuasarsSubEddington,Nesvadba11AccretionRate}
where there is no evidence for enhanced accretion after mergers
\citep{Villforth14QuasarNotMergerTriggered}.  A major factor in the
suppression in our simulations is the explicit inclusion of angular
momentum support
\citep{Hopkins11AngMomBH,Tremmel16Romulus}. In
galaxies with a dense, rotationally-supported gas disk the accretion
rates are reduced by factors up to $10$ relative to a standard
Bondi-Hoyle approach; the behaviour of the BH in different
circumstances is discussed further in Sec. \ref{sec:how-unquench}.

\subsection{Energy coupling matters more than total energy input}\label{sec:energy-coupling-matters}

\begin{figure}
\includegraphics[width=0.48\textwidth]{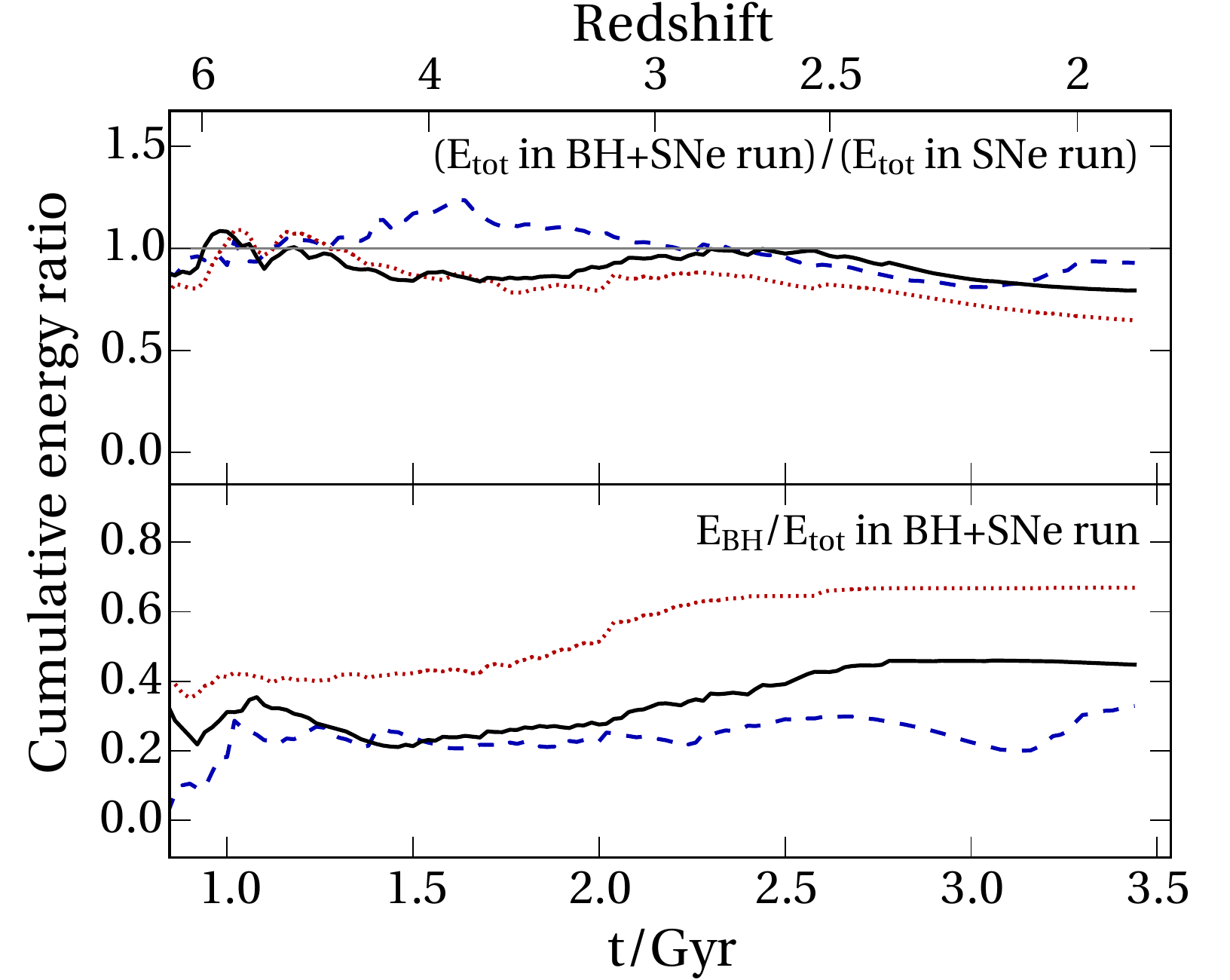}
\caption{The total feedback energy input to the galaxies is
  approximately the same in BH+SNe as in SNe-only simulations. {\it
    Upper panel}: the total cumulative energy from all feedback
  sources in the BH+SNe runs as a fraction of the total energy in the
  SNe-only runs. In all three cases, the BH+SNe runs inject comparable
  feedback energy to the SNe-only runs.  {\it Lower panel}: the
  fraction of energy delivered by the BHs in the BH+SNe suite. The BH
  energy input is significant in all cases, and dominates in the case
  of the enhanced-merger simulation. The balance between BH and SNe
  feedback determines whether large-scale outflows are generated. }\label{fig:energies}
\end{figure}

The BH feedback is essential to quenching (Fig.  \ref{fig:ssfr}); we
now consider why the supernova feedback is unable to play this role.
With our numerical parameters, the energy that couples thermally to
gas for each solar mass accreted onto the BH is
$3.6\times 10^{51}\,\ergs$, whereas the value for each solar mass of
stars formed is $6.8 \times 10^{48}\,\ergs$. Using the rates from Figs
\ref{fig:ssfr} and \ref{fig:bh-energy} respectively, we calculate the
total energy deposited by all feedback sources in the SNe-only and
BH+SNe suites. The top panel of Fig. \ref{fig:energies} shows the
energy injection accumulated over time $E_{\mathrm{tot}}(<t)$ in the
BH+SNe simulations as a fraction of that in the SNe-only
equivalents. The ratio is approximately unity, and declines over time;
in energetic terms, the reduced star formation rate in the BH+SNe
simulations slightly over-compensates for the BH energy input.

The calculation above implies that total energy input is not
the controlling factor deciding whether a galaxy quenches. Furthermore, 
\cite{Tremmel16Romulus} showed that, in test runs, doubling the total
energy input from supernovae is unable to reproduce the correct halo
mass--stellar mass relation or the sSFR as a function of
time; AGN have a qualitatively different effect to supernova feedback.

The lower panel of Fig. \ref{fig:energies} focuses on the BH+SNe
suite, showing the cumulative energy injected by BHs,
$E_{\mathrm{BH}}(<t)$, as a fraction of the total,
$E_{\mathrm{tot}}(<t)$. In the enhanced-merger case, $46\%$ of the
total feedback energy has been supplied by the BH at the time
of quenching ($1.8\,\Gyr$), rising to $63\%$ by the end of the
simulation. The corresponding figures for the reference case are 
$42\%$ (at quenching, $2.5\,\Gyr$) and $43\%$ (at the end of the
simulation, by which time star formation has resumed and the fraction
is falling). The cumulative contribution of BH feedback to the
suppressed-merger case, which never quenches, does not exceed $35\%$
at any time. 

This strongly suggests a link between quenching and the fraction of
feedback energy contributed by the AGN. To sharpen the connection,
consider the energetics of gas leaving the galaxy (enclosed
in a sphere at $10\,\kpc$ to be conservative).  We calculate, for
each particle $i$, the radial velocity $v_{r,i}$ and gas internal
energy $u_i = 3 kT_i/ 2 m_p \mu_i$ (where $k$ is the Boltzmann
constant, $m_p$ is the proton mass, $T_i$ is the temperature and
$\mu_i$ is the relative atomic mass of the gas). From here we define
the outflow specific energy summary statistic as
\begin{equation}
\textrm{Outflow energy} \equiv \frac{\sum_i m_i v_{r,i} \left( u_i
    + v_{r,i}^2/2 \right)}{\sum_i m_i v_{r,i}}\textrm{,}\label{eq:outflow-energy}
\end{equation}
where $m_i$ is the particle mass and the sum is taken over all outflowing
($v_r>0$) particles in the radial bin $9.5 < r/\kpc < 10.5$.
The expression (\ref{eq:outflow-energy}) is
constructed to give a specific energy representative of the bulk of
the outflow, using a mean weighted by local outflow rate $v_r \rho$
where $\rho$ is the local gas density.

\begin{figure}
\includegraphics[width=0.48\textwidth]{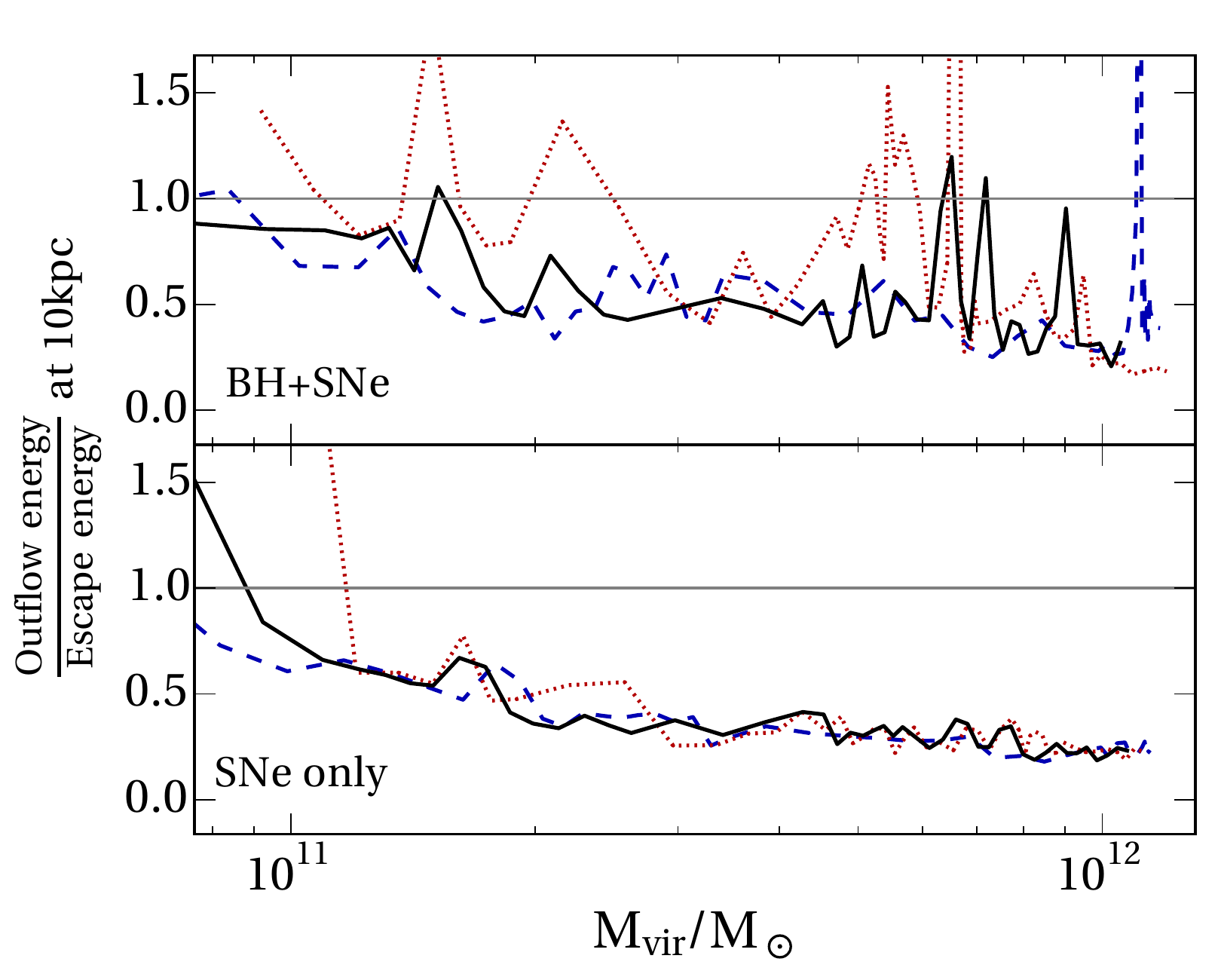}
\caption{BHs are required for outflows at dynamical masses
  $\Mvir>10^{11}\,\Msol$. The energy in outflows, defined by Eq. 
  (\ref{eq:outflow-energy}), is shown as a proportion of the energy
  necessary to escape the potential of simulated halos as they grow
  more massive over time. Independently of the assembly history,
  SN-generated winds (lower panel) are unable to leave the central
  regions of galaxies hosted in halos with mass approaching 10$^{12}$
  $\Msol$. On the other hand, BH feedback is centrally concentrated
  and therefore can create unbound winds at all halo masses (upper
  panel), even though the energy input is similar between the two
  simulation suites
  (Fig.~\ref{fig:energies}).} \label{fig:outflows-energy}
\end{figure}

Figure \ref{fig:outflows-energy} shows the outflow specific energy
defined in this way, divided by the specific energy required to
overcome the gravitational potential, defined as
$\Phi(r_{\mathrm{vir}}) - \Phi(10\,\kpc)$. The upper and lower panels
show BH+SNe and SNe-only suites. Values above unity (horizontal grey
line) indicate that the majority of the outflowing gas is able to
escape; conversely, values below unity imply that only a small
proportion, if any, of the gas is able to escape the potential.  We
choose to plot against the virial mass of the halo (which
monotonically increases with time) because, in the case of supernovae,
the outflow properties are then expected to be approximately
independent of merger history \citep{KellerWadsley16}.

The BH+SNe and SNe-only results are starkly different despite the two
simulation suites invoking similar total feedback energy inputs
(Fig. \ref{fig:energies}).  The lower panel of
Fig. \ref{fig:outflows-energy} shows that supernova-driven outflows
are increasingly ineffective at leaving the halo as the mass
increases. This is consistent with the results of
\cite{Christensen15outflows} who analysed the same SF prescription and
found a mass outflow rate declining steeply with halo circular
velocity. Conversely, the upper panel shows that BH feedback is able
to eject gas from the halo at any mass.

We are now in a position to understand why BHs are essential to
quenching. The BH feedback couples a significant fraction of the total
available energy to a small mass at the very centre of the disk. This
generates temperatures of order $10^7\,\K$ on the smallest resolved
scales of a few hundred parsecs, while supernova bubbles rarely exceed
$10^6\,\K$. As the over-pressurised gas expands, the BHs give rise to
high velocities but low gas densities in the centre of the galaxy;
we find that typical mass outflow rates from the central kiloparsec
are around $2$ to $5$ times {\it lower} in the BH+SNe suite.

There are two reasons why this low-mass, high-temperature feedback is
more effective than supernovae by the time that we measure the outflow
leaving the galaxy at $10\,\kpc$. First, the cooling time is longer;
second, the total gravitational energy penalty for leaving the disk is
lower. Both effects leave more of the feedback energy available for
coupling to the halo gas. In the SNe-only suite, the mass outflow
rates decline with radius as the gravitational and radiative energy
losses cause the outflow to stall. Conversely in the BH+SNe cases, the
mass outflow rates rise with radius because the initial energy losses
are small and mass is swept up in an expanding shocked shell. It is
this large-scale outflow, alongside the associated halo gas heating,
that initiates and sustains galaxy quenching, as we now discuss.

\subsection{The AGN is required to actively sustain quenching}\label{sec:sustain-quenching}

\begin{figure}
\includegraphics[width=0.48\textwidth]{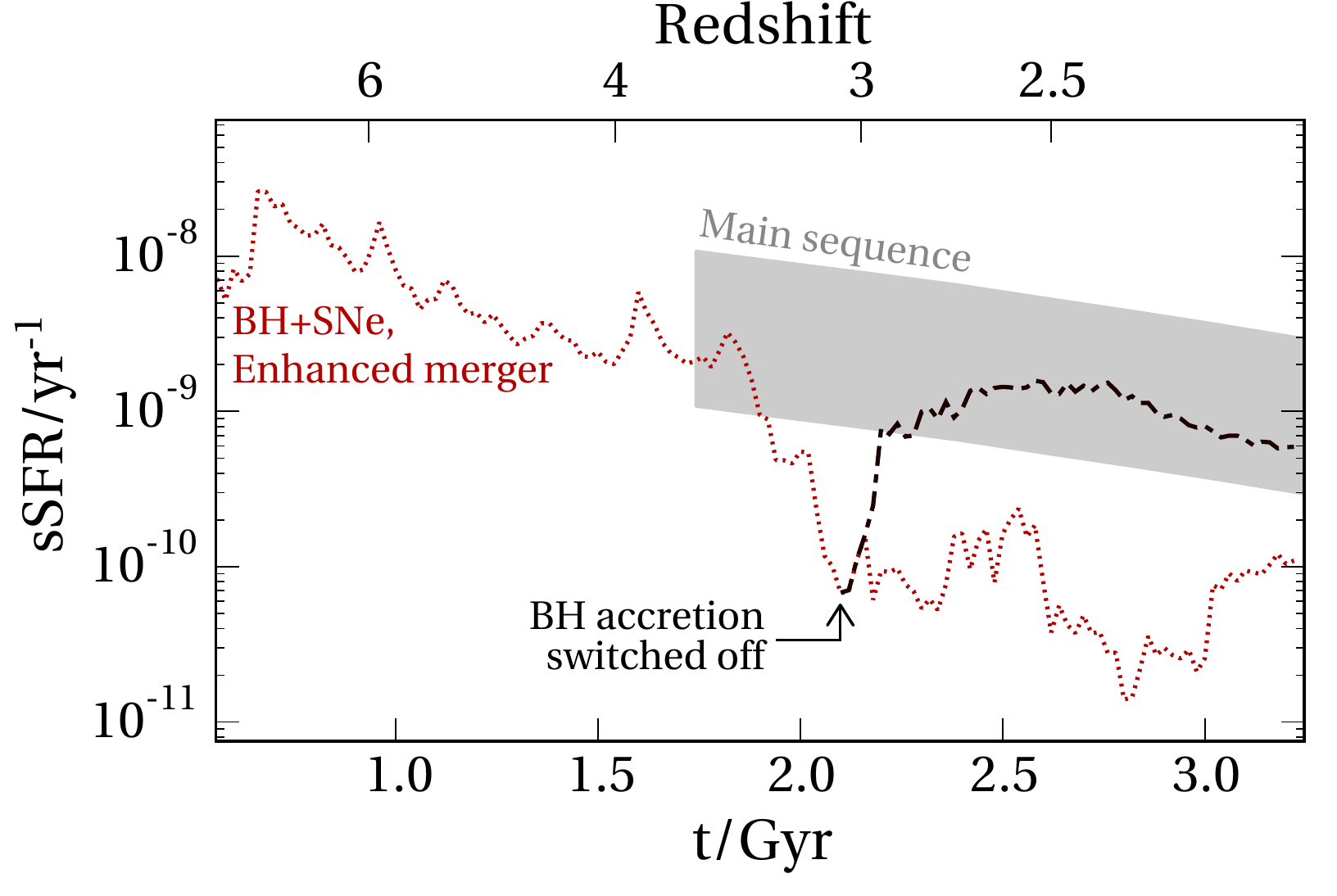}
\caption{The BH is able to suppress star formation for periods much
  longer than the central cooling time ($100\,\Myr$).  As a test, we
  restarted the enhanced-merger BH+SNe simulation from $z=3$
  ($t=2.17\,\Gyr)$ with BH accretion artificially switched off,
  finding that the galaxy rapidly rejoins the main sequence by
  reforming a star-forming disk (black dash-dotted
  line). } \label{fig:reignition}
\end{figure}

The enhanced-merger BH+SNe simulation quenches for at least
$1.5\,\Gyr$, to the end of the simulation. During this time, star
formation rate drops by two orders of magnitude (Fig. \ref{fig:ssfr})
but the BH accretion rate remains steady
(Fig. \ref{fig:bh-energy}). Since the cooling time in the inner
$10\,\kpc$ of the halo is only $100\,\Myr$, this continuing feedback
is crucial to maintaining the quenching
\citep[e.g.][]{Croton06,Sijacki07AGNfeedback}. We directly verified
that the BH energy input is required by restarting our enhanced-merger
BH+SNe simulation from $z=3.0$ ($t=2.17\,\Gyr$), manually turning off
the BHs after the sSFR has dropped to below $0.1 \, \Gyr^{-1}$. The
results are shown in Fig.  \ref{fig:reignition}; star formation is
rapidly re-established and the galaxy rejoins the main sequence in
less than $200\,\Myr$. This underscores the importance of BHs to
maintaining (as well as establishing) the quenched state.

\begin{figure}
\includegraphics[width=0.48\textwidth]{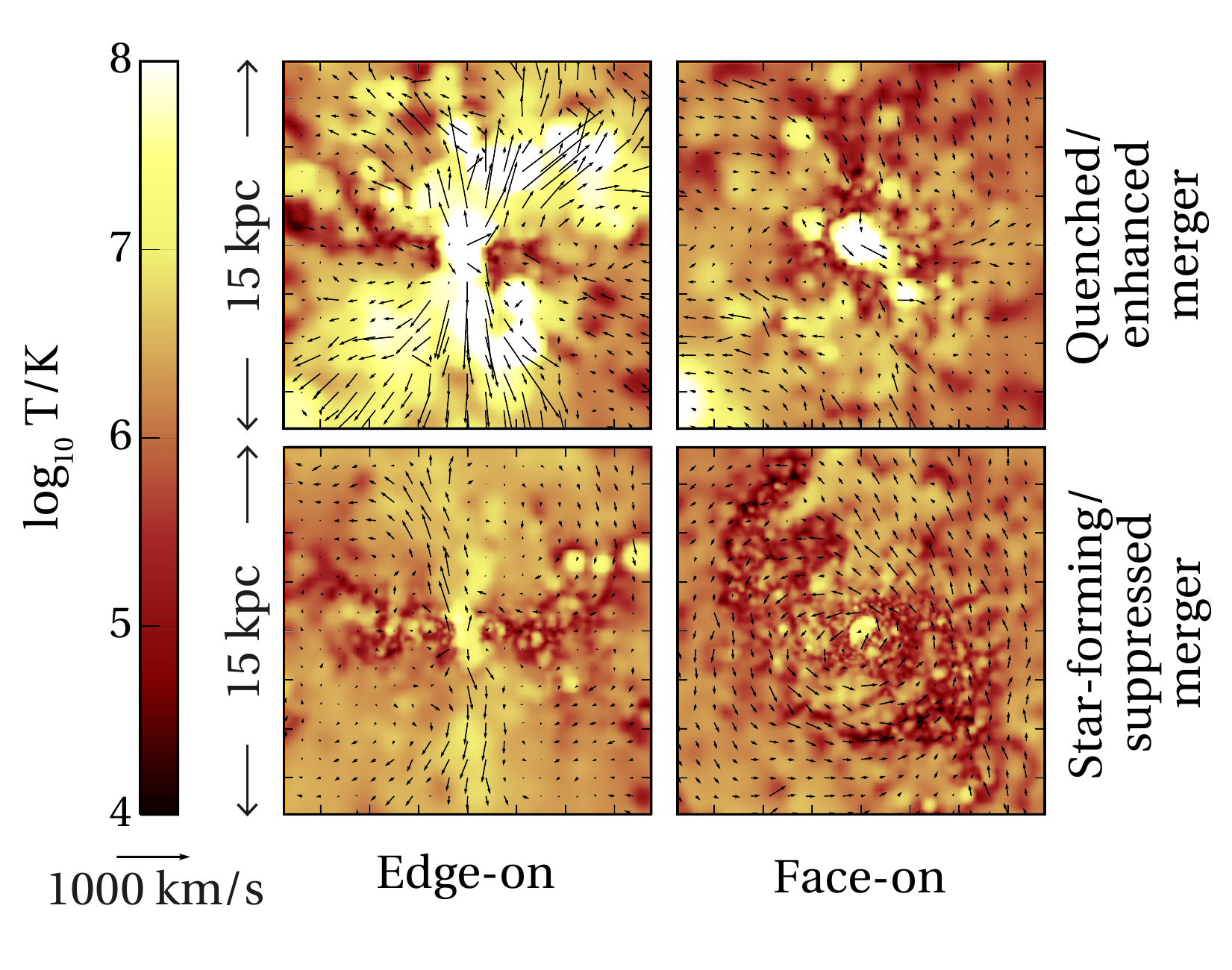}
\caption{Temperature maps of the central $15\,\kpc$ show how the
  effect of the BH is strongly dependent on the gas around it. {\it
    Upper panels}: the enhanced-merger simulation shortly after
  quenching. The rotation-supported disk has been destroyed, replaced
  by a turbulent environment which can rapidly remove angular momentum
  from incoming cool material, maintaining BH accretion. {\it Lower
    panels:} The suppressed-merger simulation at the time of maximum
  star formation rate. Tidal forces have compressed the disk, but the
  angular momentum support restricts accretion onto the BH. Any energy
  released from the BH is strongly confined by the surrounding dense
  gas, so that it exits perpendicular to the disk and does not disrupt
  ongoing star formation. }\label{fig:bh-accretion-mode}
\end{figure}

It is worth investigating why the BH continues to accrete even when
star formation cannot proceed, despite the two processes sharing a
common fuel in the form of cooling gas.  Figure
\ref{fig:bh-accretion-mode} shows temperature maps of the gas in (top
row) the enhanced and (bottom row) the suppressed-merger BH+SNe
simulations. The left and right panels of each row show edge-on and
face-on slices, as defined by the stellar angular momentum. First
consider the quenched galaxy (top row). Crucially, there is cool
($\sim 10^4\,\K$) gas present which is able to feed the BH. Within
$5\,\kpc$ of the centre the total mass of neutral and molecular gas
(which we do not distinguish in these simulations) is
$1.2 \times 10^8\,\Msol$, compared to $4.0 \times 10^9\,\Msol$ in the
SNe-only run. Observations indeed indicate that early type galaxies
contain cold gas reservoirs in agreement with this result
\citep[e.g.][]{Davis14ColdGasInEarlyTypes}. The primary source of cool
gas in both simulations is infalling streams.

When quenched, the turbulent remnant of the disk is able to rapidly
remove angular momentum from new material. As a result, the greatest
BH accretion rates are reached somewhat after the galaxy has
quenched; the initial event involves mechanically disrupting the
ordered rotation of the disk, after which the BH is able to
cause substantial damage to the remaining cool interstellar
medium. Observations show that BH activity can continue after
star formation is shut off \citep[e.g.][]{Nandra07AGNhostgalaxies}, in
agreement with this picture. Cool material that does not
reach the central BH is typically disrupted by the rapid hot
outflows in the vicinity.

Conversely in the star-forming mode (bottom row), infalling gas joins
the ordered rotation of the disk, minimising the angular momentum
losses. Once in a long-term stable orbit, the new gas fuels star
formation. Our implementation of BH accretion includes the physical
support provided by angular momentum on resolved scales
\citep{Tremmel16Romulus}, which limits the accretion rate even when
the BH is surrounded by dense gas. Furthermore the effects of the
feedback are confined by the presence of the disk, meaning that the
star-forming regions are essentially shielded from any heating; the
AGN energy is instead directed into the halo, perpendicular to the
disk and cool gas inflow. This contrasts with the quenched
system in which lack of pressure confinement for the BH feedback
allows it to indefinitely prevent the re-formation of the disk.

The reference run serves as a helpful intermediate case where the
cycle of quenching described above terminates (see Fig. \ref{fig:ssfr}): star formation resumes
because the outer halo cools and provides
enough dense gas to overwhelm the AGN feedback cycle. We investigate this
more thoroughly below, after discussing how the suppressed-merger run
avoids quenching entirely.

\subsection{AGN activity can coexist with star formation}\label{sec:how-unquench}

The suppressed-merger simulation never shuts off star formation; the
rate peaks at $100\,\Msol\,\yr^{-1}$ at $t=3.1\,\Gyr$ and gently
declines to $20\,\Msol\,\yr^{-1}$ at $t=3.5\,\Gyr$ (by which time it
has a total stellar mass of $5.3\times 10^{10}\,\Msol$). We verified
that, extending the simulation another timestep to $t=3.6\,\Gyr$, 
the galaxy remains on the star-forming main-sequence.

The peak in star formation rate is caused by the encounter with system
E (Fig. \ref{fig:portraits}). Closest approach occurs at
$t=2.8\,\Gyr$; the tidal forces compress the gas disk and enhance star
formation in both SNe and BH+SNe runs (Fig. \ref{fig:ssfr}). The
enhanced, centrally concentrated star formation causes oscillation
within the main sequence
\citep{Zolotov15Quenching,Tacchella16Compaction}. As previously stated
in Sec. \ref{sec:how-quench}, BH accretion rates also increase
immediately following the tidal encounter, which drives a significant
outflow for the first time in the history of the suppressed-merger
galaxy. However the outflow is directed perpendicular to the disk
which mitigates its impact on the instantaneous star formation rates,
and the galaxy continues to be fuelled with fresh gas from the
intergalactic medium (Fig. \ref{fig:inflow-10kpc}). The galaxy
therefore does not quench; luminous AGN, star formation and rapid
outflows can all coexist in the same galaxy, as seen in observations
\citep{Nandra07AGNhostgalaxies,ForsterSchreiber14SinsOutflows,Carniani16QuasarsDontAlwaysQuench}.

\subsection{The longevity of quenching depends on merger ratio}\label{sec:temporary-or-permanent}

As discussed in Sec. \ref{sec:sustain-quenching}, the
enhanced-merger simulation with BH+SNe feedback quenches from shortly
after the galaxies merge until the end of the simulation.  However,
the reference simulation quenches temporarily before reforming a
star-forming disk. It rejoins the main sequence $1.5\,\Gyr$ after
leaving it (Fig. \ref{fig:ssfr}); by the more stringent UVJ cut,
the total period for which it appears quenched is less than
$1\,\Gyr$. At the end of the simulation it has a total stellar mass of
$3.2\times 10^{10}\,\Msol$ and is forming new stars at a rate of
$16\,\Msol\,\yr^{-1}$.

\begin{figure}
\includegraphics[width=0.48\textwidth]{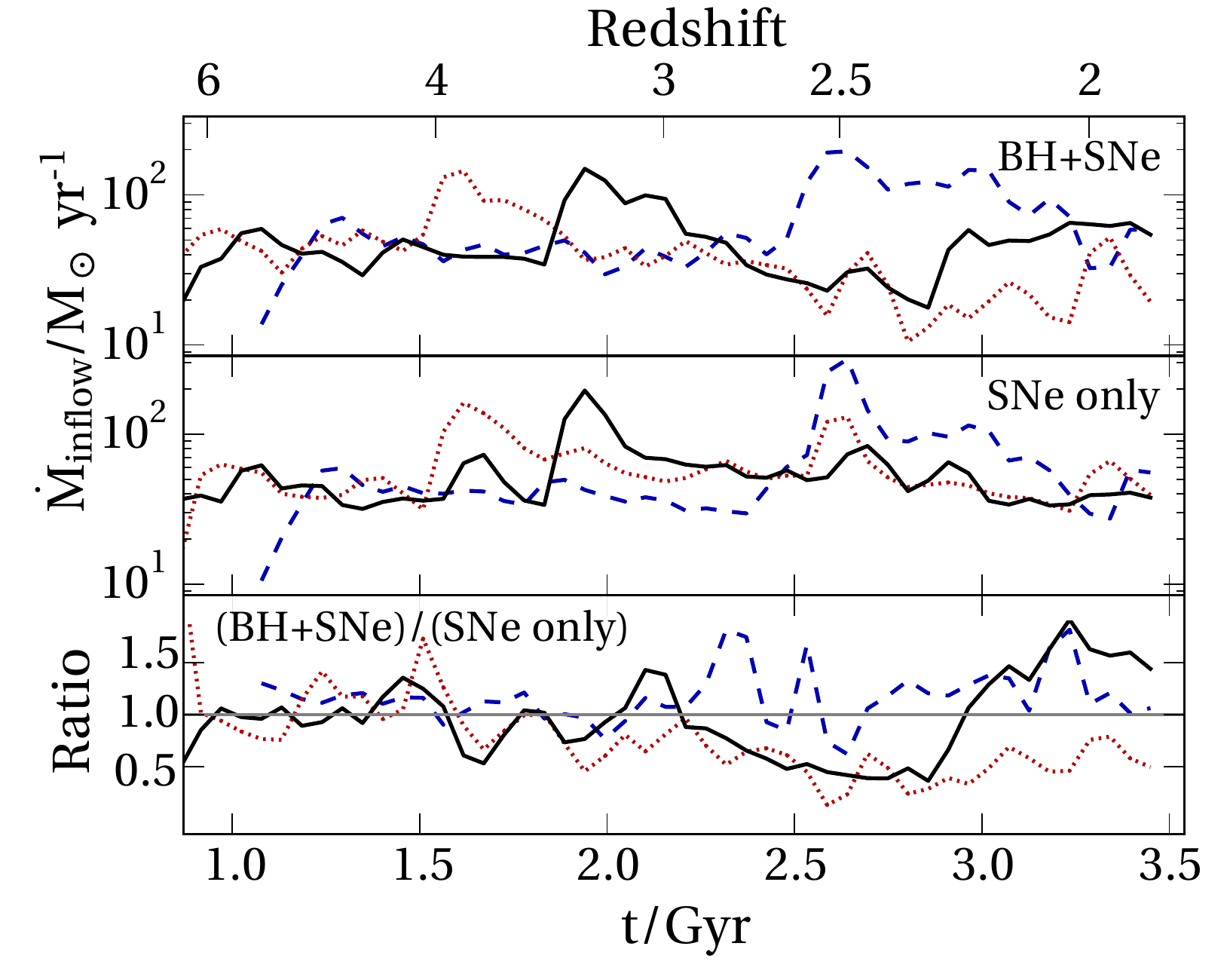}
\caption{Long-term quenching requires the galaxy inflow to be
  suppressed. {\it Upper two panels:} The rate at
  which gas is inflowing across a fixed radius of $10\,\kpc$ for
  BH+SNe and SNe-only suites respectively. {\it Bottom panel:} The
  ratio between the inflow in the two suites; accretion onto the
  galaxy is strongly suppressed for long periods by the BHs in the
  enhanced-merger case (red dotted line), but recovers at
  $t = 3\,\Gyr$ in the reference case (black solid line). In the
  reference case, this leads to the reformation of the disk and
  readmission of the galaxy onto the star-forming main sequence,
  whereas the enhanced-merger galaxy never recovers
  (Fig.~\ref{fig:ssfr}).} \label{fig:inflow-10kpc}
\end{figure}

The difference between the two cases is understood by inspecting the
rate of gas accretion into the galaxy (measured on a sphere at
$10\,\kpc$ from the halo centre) as depicted in
Fig. \ref{fig:inflow-10kpc}. The top two panels show the rate in
SNe-only and BH+SNe suites respectively, with the bottom panel showing
the ratio. In the reference {and enhanced} scenarios, BH
feedback has the effect of reducing the accretion rate by up to a
factor of two when the galaxies are quenched at
$t=2.5\,\Gyr$. {Conversely in the suppressed scenario, inflow
  rates are only weakly dependent on feedback
  -- and, counterintuitively, the average inflow rate is slightly
  higher in the BH+SNe case because some of the outflowing material is recycled in
  a galactic fountain.}  In the reference case, this fountain effect
is seen after $3.0\,\Gyr$: {inflow compensates for the
  earlier BH-induced outflows. Such a compensation never occurs in the
  enhanced case over the lifetime of our simulation.}

{ The three scenarios illustrate how the inflow
  rate depends critically on whether previously expelled material is
  able to cool or not.}
In the enhanced-merger case, the mean temperature of the circumgalactic
medium at $150\,\kpc$ (measured at $t=2.8\,\Gyr$) is
$T=2\times 10^5\,\K$ in the enhanced-merger case, as opposed to
$T=0.8\times 10^5\,\K$ in the reference and suppressed-merger
cases. This results in a cooling time at these radii of around
$5\,\Gyr$ for the enhanced-merger case, contrasting with $1\,\Gyr$ in
the other cases. Additionally, the cool inflowing material is easily
disrupted in the more turbulent environment of the enhanced-merger
simulation.

In summary, the reference simulation is able to regrow a disk because
accretion continues and overwhelms the ability of the BH to
expel material; conversely the enhanced-merger simulation maintains a
lower accretion rate due to the higher initial energy input into the
halo, uses the low level of accretion to directly fuel the BH,
and consequently does not regrow its disk.

\subsection{The fraction of quenched galaxies can be predicted}\label{sec:semi-enhanced-sim}

The discussion above explains how the enhanced-merger simulation, with
a ratio of 2:3, quenches indefinitely whereas the reference (1:5)
simulation quenches for $\sim 1\,\Gyr$. In an attempt to refine the
merger ratio required for long-term quenching, we ran an additional
BH+SNe simulation that is intermediate between the reference and
enhanced cases. We refer to the additional simulation as the {\it
  semi-enhanced} run. The ability to generate new cases to test
hypotheses in this way is a major benefit of the GM
approach (Sec. \ref{sec:genet-modif-appr}).

\begin{figure}
\includegraphics[width=0.48\textwidth]{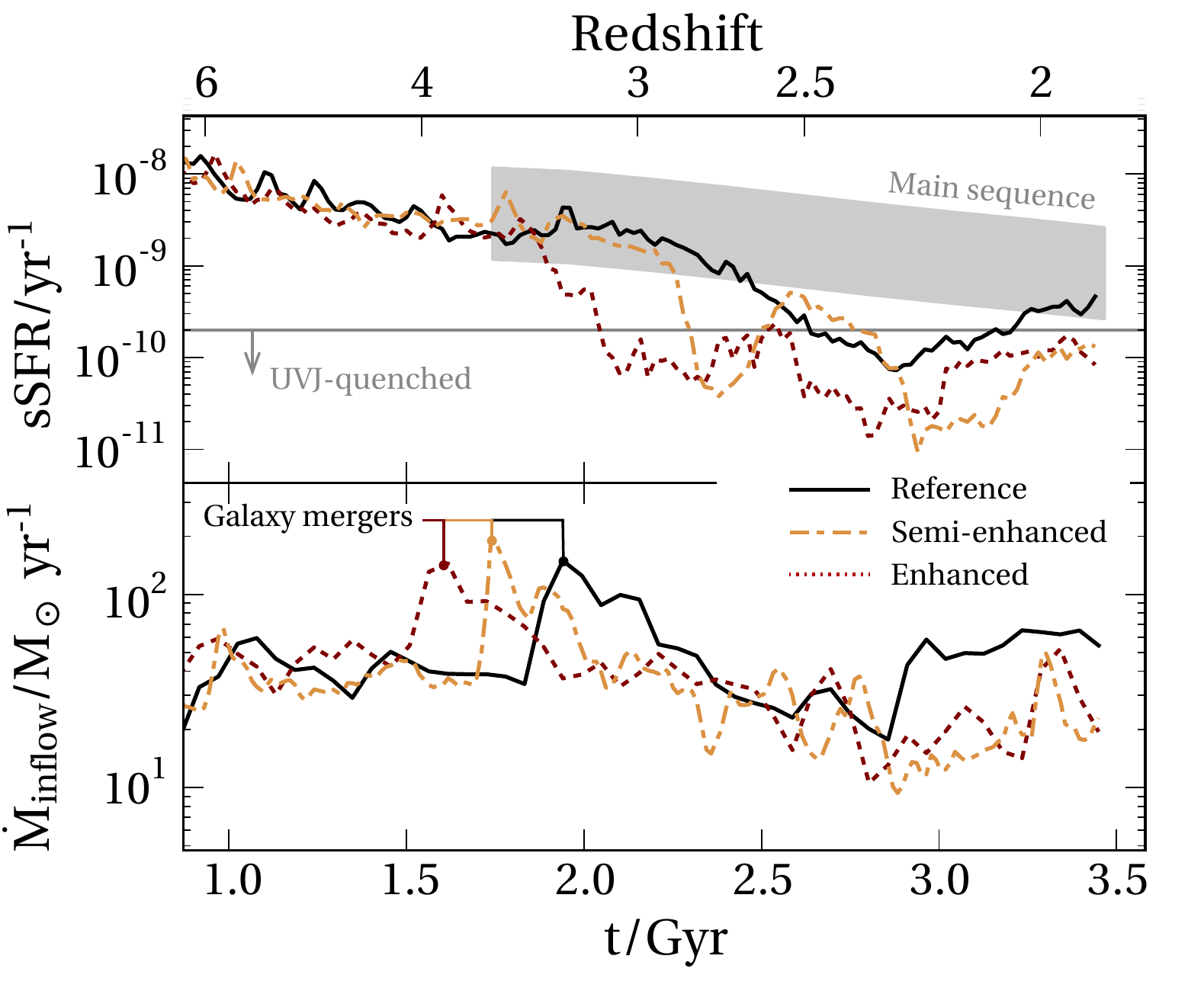}
\caption{Using the genetic modification approach, we 
  generate a semi-enhanced scenario in which the history is
  intermediate between the reference and enhanced-merger cases.  The
  resulting galaxy has a 1:3 merger (as opposed to 2:3 and 1:5 for the
  enhanced and reference scenarios). {\it Upper panel}: The specific
  star formation rate of the new scenario with BH+SNe feedback (orange
  dash-dotted line). The enhanced (red dotted) and reference (black
  solid) cases are also plotted. The onset of quenching in the
  semi-enhanced case is intermediate between the two.  Subsequently a
  fountain leads to some of the original disk reforming, although it
  is rapidly depleted. {\it Lower panel}: the inflow rate at 10 kpc
  for the three simulations. The times of the respective galaxy
  mergers are indicated. The semi-enhanced case suppresses inflow (and
  therefore long-term star formation) in the same manner as the
  enhanced simulation.} \label{fig:semi-enhanced-ssfr}
\end{figure}

Figure \ref{fig:semi-enhanced-ssfr} shows results from the
semi-enhanced simulation (orange dash-dotted lines) alongside the
reference and enhanced counterparts (solid black and dotted red
lines). The halo merger in the semi-enhanced run takes place at
$t=1.6\,\Gyr$ with a ratio of 1:3; the galaxies merge at $t=1.8\,\Gyr$
($z=3.5$). In the top panel, we show the specific star formation rate;
like the enhanced case, the semi-enhanced run quenches shortly after
its merger. The galaxy then weakly recovers its star formation between
$t = 2.5\,\Gyr$ and $2.7\,\Gyr$. This transitory effect results from
a fountain-like recycling of the cool galaxy disk remnants. However
the long-term galaxy accretion rate (lower panel) follows the
enhanced-merger galaxy and so the star formation cannot be sustained.

Using the behaviour as a function of merger ratio, and tentatively
adopting the assumption that this is the only variable determining
whether a galaxy with a BH shuts off star formation or not, we can
estimate the size of the quenched population. We take all halos from
our dark-matter-only $(50\,\Mpc)^3$ simulation (Sec.
\ref{sec:simulations}) at $z=1.8$ in the mass range
$0.8 \times 10^{12} < M_{\mathrm{vir}}/\Msol < 1.2 \times 10^{12}$ (a
total of 57 objects). We construct merger trees and find the largest
merger event on the major progenitor branch for each of these halos up
to $z=4$, to match the merger events considered in this paper. Six halos
($11\%$) have a merger at least as significant as 2:3; 15 ($26\%$)
more significant than 1:3 and 31 ($54\%$) more significant than 1:5.

In a survey at $z \simeq 1.8$, one can therefore expect roughly a
quarter of all objects at $10^{12}\,\Msol$ (or stellar masses of
$\simeq 3 \times 10^{10}\,\Msol$) to be fully quenched, in agreement
with {observational estimates from a study of CANDELS 3D-HST
  fields} \cite[][]{Lang14Candels3dHst}.  This picture
predicts that a further quarter of objects in such a survey will have
quenched at some point in their past but subsequently ``rejuvenated''
and returned to the main sequence. 

{An attraction of working with the GM approach is
  that one can isolate different variables controlling a galaxy's
  history and environment (see Sec. \ref{sec:concl-disc}). Here we
  have worked on the hypothesis that major merger ratio is the most
  significant parameter; in future work we will discuss whether the
  predictions for quenched fractions remain robust when other aspects
  of a galaxy are allowed to change.}

\subsection{AGN activity prevents over-contraction of the dark halo}\label{sec:dynamical-effects}

\begin{figure}
\includegraphics[width=0.48\textwidth]{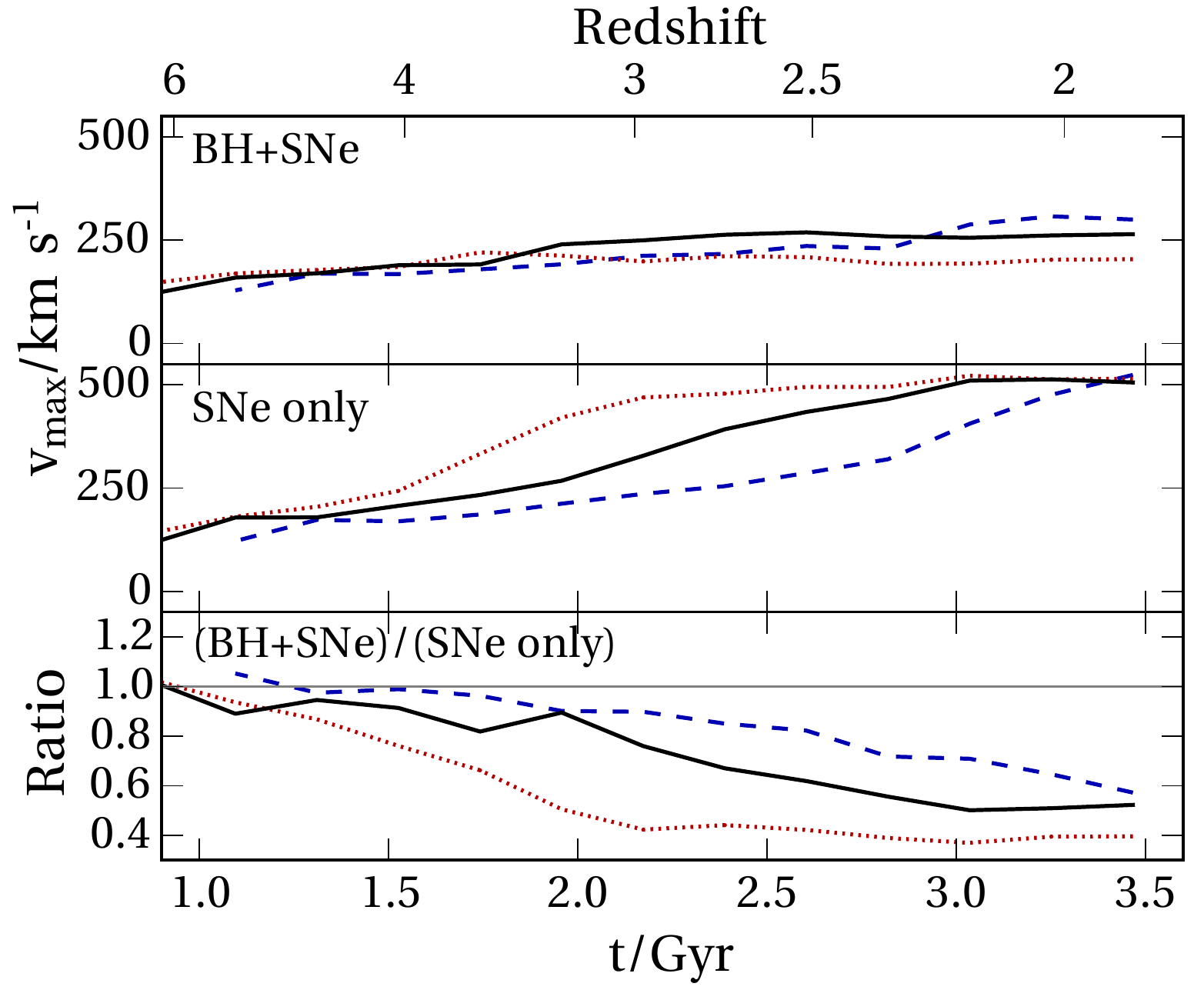}
\caption{BHs regulate the build-up of baryonic mass in the centre of
  the galaxy. {\it Upper two panels:} The central concentration of
  baryons is quantified by measuring $\vmax$, the peak value of the
  rotation curve in the disk plane, for BH+SNe and SNe-only suites
  respectively. In the SNe-only case, $\vmax$ continues to grow as
  mass accretes and falls into the centre, forming a bright central
  bar. In the BH+SNe case, a limit is reached at
  $\vmax \simeq 250 \,\kms$.  {\it Bottom panel:} The ratio of the
  simulations shows how AGN feedback at first has little effect on the
  dynamics, but as the BH mass grows, it becomes able to eject low
  angular-momentum material, preventing excessive central star
  formation. } \label{fig:vmax}
\end{figure}

The ability of BHs to eject baryons from deep potential wells (Sec.
\ref{sec:how-quench}) has knock-on effects on the overall dynamics of
galaxies. The most direct consequence is that the central
concentration of baryons is greatly reduced, changing the rotation
curve. Figure \ref{fig:vmax} shows the maximum circular velocity speed
$\vmax$ (measured in the disk plane) as a function of time in each
simulation. The BH+SNe suite forms galaxies with
$v_{\mathrm{max}}\simeq 250\,\kms$ (top panel) whereas in the SNe-only
suite (bottom panel), all galaxies reach $v_{\mathrm{max}}= 500\,\kms$
which is unrealistically high \citep[e.g.][]{Wisnioski15Kmos3d}. The
build-up of $\vmax$ over time in this SNe-only case closely tracks the
gas accretion rate into the halo. When BH feedback is active, however,
low angular momentum gas reaching the centre is efficiently ejected,
so that growth of the galaxy's $\vmax$ decouples from the total
accretion even in the suppressed-merger case where the BH has
relatively little impact on the overall star formation rate.
AGN have also been shown to be critical for preventing over-contraction
of early-type galaxies at $z=0$ \changed{\citep{Martizzi12,Dubois13AGNGalFormation}}.

The dynamical time at the centre of our BH+SNe galaxies is
approximately $100\,\Myr$. Figure \ref{fig:bh-energy} shows that the
BH accretion rate, and therefore energy input, varies on
timescales considerably shorter than this period. According to the
analytic arguments of \cite{PG12}, this should result in the expulsion
of dark matter from the centre of the galaxy, and ultimately to the
formation of a near-constant-density dark matter core \cite[see
also][]{Martizzi12,PG14}. For example, in the enhanced merger BH+SNe
simulation, the limiting dark matter density profile (fitted in the
innermost kiloparsec at $t=3.5\,\Gyr$) follows $\rho \propto
r^{-0.7}$, which is significantly shallower than the $\sim r^{-1}$
scaling of a pure dark matter simulation. Conversely in the
enhanced-merger SNe-only simulation, the same measurement gives $\rho
\propto r^{-2.0}$, indicating a very strong adiabatic contraction from
the dense pile-up of baryons. Our simulated BHs are therefore 
crucial in controlling not just the rate of star formation but 
the overall distribution of matter in the centre of galaxies.

The accretion of baryons and dark matter onto the halo is also
affected by the BH feedback. The baryon fractions in the SNe suite
remain at the cosmic mean fraction throughout the simulations, but
strongly decline in the BH+SNe cases to $62\%$, $87\%$ and $95\%$
respectively\footnote{{However at the time of the most major
    merger, these differences are much smaller. The gas fraction of
    the infalling structure is $104\%$, $99.7\%$ and $96.2\%$
    ($105\%$, $103\%$ and $104\%$) of the cosmic mean value in
    respectively the BH+SNe (and SNe-only) suite for suppressed,
    reference and enhanced cases. }}. This should have observable
effects on the density of absorbers around early- and late-type
galaxies \citep{Suresh15IllustrisBimodalOVIAbundance}. The dynamics of
the dark matter accretion are also affected, since the reduction in
gravitational mass slows subsequent accretion
\citep{Schaller15BaryonsAndDmEagle}. As only the BHs drive dynamically
significant outflows, the BH+SNe galaxy halos have final dark matter
masses that are $88\%$, $95\%$ and $99.5\%$ of the SNe-only masses for
the enhanced, reference and suppressed-merger cases respectively.

\section{Conclusions}\label{sec:concl-disc}

In this paper, we investigated how galaxy mergers and black hole (BH)
feedback work together to quench star formation. We used ``genetically
modified'' initial conditions to set up three galaxies which live in
the same environment and large scale structure, but that differ in
merger history. Coupled to a new implementation of BH physics
\citep{Tremmel16Romulus}, we were able to isolate how mergers lead to
quenching.

Our results show that quenching in field galaxies with virial masses
larger than a few times $10^{11}\,\Msol$ can only be achieved using BH
feedback (Fig. \ref{fig:ssfr}). Supernova feedback becomes ineffective
at driving galactic winds at these masses
\citep{Christensen15outflows,KellerWadsley16}. However the addition of
BH feedback does not increase the total energy budget. Instead, its
strong central concentration generates low-density outflows at high
temperatures and velocities, minimising radiative energy losses. As
the wind moves outwards, it sweeps up halo material, increasing the
outflow density only after the deepest part of the gravitational
potential has already been overcome. In this way the total energy loss
due to purely gravitational effects is minimised.

Mergers trigger quenching without invoking a near-Eddington quasar
phase for the central AGN. The average BH accretion rate agrees with
constraints from stacking \citep{Mullaney2012}. All of our galaxies,
regardless of the merger history, have an active BH at their
centre, in keeping with observed AGN in both quenched and
unquenched systems \citep{Simmons12AGN, Mullaney2012, Rosario15,
  Mancini15}, driving a large-scale galactic wind \citep{Genzel14,
  Harrison14,ForsterSchreiber14SinsOutflows}.  The cold disk of gas in
the unquenched galaxies slows BH accretion due to angular
momentum support.  It also confines the effect of the BHs and
directs it outwards in a funnel, allowing star formation to proceed
despite the rapid central outflows
\citep{CanoDiaz12QuasarQuenchingPlusSFElsewhere,Carniani16QuasarsDontAlwaysQuench}. 

Mergers start the quenching process through mechanical disruption of
the cold disk. Subsequently the AGN feedback is able to have far
greater impact in disrupting the existing star-forming gas and cool
gas inflows.  Acting together in this way, the merger and central BH
push the galaxy into a self-sustaining, long-term quiescent state. The
drop-off from the main sequence is rapid, occurring over around
$250\,\Myr$ in both our reference and enhanced-merger
simulations. Many clues suggest that quenching is indeed rapid in real
galaxies \citep{Barro13,Barro15,Mancini15}. In the simulations,
quenching is followed by a slow decline in BH activity 0.5 -- 1 Gyr
after the galaxy leaves the main sequence, in agreement with
observational evidence that AGN in star forming galaxies contribute
the majority of the X-ray luminosity density \citep{Mancini15,
  Rodighiero15}.  The quenching mode is then maintained by the
turbulent remnants of the disk sweeping a low level of infalling cool
gas into the BH, which allows BH accretion to continue as
seen in recent observations by \cite{Tremblay16AGNColdAccretion}. One
prediction of this scenario is that quiescent galaxies will have an
offset BH-stellar mass relation compared to star-forming galaxies
\citep{Terrazas16SemiAnalyticAGNQuenching}.

The BH feedback has a significant dynamical impact on the galaxy. Gas
that is accreted into the disk at early times tends to carry little
angular momentum; in the absence of a strong outflow or cycling
mechanism, this piles up, creating a strong central bar
(Fig. \ref{fig:portraits}, third row of panels). The maximum circular
speed in the disk plane reaches $\vmax \simeq 500\,\kms$ in all three
merger scenarios when only stellar feedback is available
(Fig. \ref{fig:vmax}, top panel). The stronger outflows of the BH+SNe
simulations prevent this pile-up from occurring, using a combination
of a galactic fountain and outright ejection to set a limit of
$\vmax \simeq 250 \kms$. {There is considerable interest in
  integral field spectroscopy of high-redshift galaxies
  \citep[e.g.][]{ForsterSchreiber06SINS,Newman15,Wisnioski15Kmos3d},}
and we anticipate that the kinematics of galaxies can therefore be
used as an indicator for the balance between BH and stellar feedback.

Our study hints at the broader potential of the ``genetic
modification'' (GM) technique \citep{Roth15GM} to shed light on
problems in galaxy formation. Because our three scenarios (reference,
enhanced and suppressed) share the same environment and large scale
structure (Fig. \ref{fig:portraits}), with the central galactic disk
assembling from gas streaming along the same filaments, we have been
able to construct a clean test of the effect of merger ratio. The GM
approach can search systematically over histories, {providing
  controlled tests which retain the benefits of idealised simulations
  in pinning down the effect of merger ratio
  \cite[e.g.][]{Johansson09} while also including cosmological
  accretion}. We illustrated this point by constructing a scenario
with a merger ratio intermediate between the reference and enhanced
cases (Fig. \ref{fig:semi-enhanced-ssfr}). The results allowed us to
estimate that a quarter of systems in our mass range at $z\simeq 2$
should appear quenched, and a further quarter will have recovered from
an earlier episode of quenching -- assuming that merger ratio is the
significant variable controlling the outcome. Future work could extend
this study by changing, for example, the mean density of the region
within which the galaxy is hosted to directly interrogate the
important role of environment
\citep[e.g.][]{Peng10Quenching,Wijesinghe12gamaEnvironmentSFR}.

\section*{Acknowledgments}
\changed{We are grateful to the anonymous referee for helpful comments}. All
simulation analysis made use of the \textsc{pynbody}
\citep{2013ascl.soft05002P} and \textsc{tangos} (Pontzen et al. in
prep) suites. AP and AS were supported by the Royal Society. MT, FG
and TQ were partially supported by NSF award AST-1514868.  NR and HVP
were partially supported by the European Research Council under the
European Community's Seventh Framework Programme (FP7/2007-2013) / ERC
grant agreement no 306478-CosmicDawn. NR was additionally supported by
STFC. MV acknowledges funding from the European Research Council under
the European Community's Seventh Framework Programme (FP7/2007-2013
Grant Agreement no. 614199, project ``BLACK'').  FG was partially
supported by NSF AST-1410012, HST AR-14281 and NASA NNX15AB17G
grants. This work used the DiRAC Complexity system, operated by the
University of Leicester IT Services, which forms part of the STFC
DiRAC HPC Facility (\url{www.dirac.ac.uk}). This equipment is funded
by BIS National E-Infrastructure capital grant ST/K000373/1 and STFC
DiRAC Operations grant ST/K0003259/1. DiRAC is part of the National
E-Infrastructure.

\bibliographystyle{mn2e} 
\bibliography{./refs}

\end{document}